# Resonant Meta-atoms with Nonlinearities on Demand

*Dmitry Filonov[1,2,\*], Yotam Kramer[1,\*], Vitali Kozlov[1], Boris A. Malomed[1,2], and Pavel Ginzburg[1,2,\*\*]*

[1]School of Electrical Engineering, Tel Aviv University, Tel Aviv, 69978, Israel
[2]ITMO University, St. Petersburg 197101, Russia
*equal contribution
*\*pgiznburg@post.tau.ac.il*

**Abstract:** Nonlinear light-matter interactions and their applications are constrained by properties of available materials. The use of metamaterials opens the way to achieve precise control over electromagnetic properties at a microscopic level, providing new tools for experimental studies of complex nonlinear phenomena in photonics. Here a doubly resonant nonlinear meta-atom is proposed, analyzed and characterized in the GHz spectral range. The underlying structure is composed of a pair of split rings, resonant at both fundamental and nonlinear frequencies. The rings share a varactor diode, which serves as a microscopic source of nonlinearity. Flexible control over the coupling and near- and far-field patterns are reported, favoring the doubly resonant structure over other realizations. Relative efficiencies of the second and third harmonics, generated by the diode, are tailored by dint of the double-ring geometry, providing a guideline for selecting one frequency against another, using the design of the auxiliary structures. The on-demand control over the microscopic nonlinear properties enables developing a toolbox for experimental emulation of complex nonlinear phenomena.

**Introduction** Nonlinear systems feature a variety of phenomena of great significance for both fundamental studies and applications. Being often challenging for detailed mathematical analysis, they may be studied by applying experimental tools and cross-disciplinary concepts. In particular, nonlinear optics, being a celebrated topic by itself[1], also provides tools for the emulation of cosmological effects[2], among many others. Nonlinear optical interactions, in the majority of cases, require the use of high light intensities, due to naturally small susceptibilities of available materials. As a result, observation of many effects is extremely challenging and requires highly sophisticated equipment. The concept of metamaterials, however, provides guidelines for constructing artificial media with novel electromagnetic properties[3]. Careful design of resonant characteristics of the unit cells forming a composite material enables tailoring its linear and nonlinear responses alike[4]. Because efficiencies of nonlinear interactions depend on the power of a local electromagnetic field, its enhancement by means of auxiliary structures is beneficial. Various configurations in both optical and radio-frequency (RF) domains were proposed, demonstrating dramatically improved nonlinear responses of the hybrid systems (see review[5]). Nonlinear-conversion efficiencies may be further improved by employing doubly resonant structures, which concentrate near fields at both the fundamental and multiple (nonlinear) frequencies[6,7,8,9]. This cascaded enhancement is based on recycling both the pump and the nonlinear fields in a cavity, where the latter may also originate from spontaneous vacuum fluctuations [10]. Hybrid optical sources with tailored nonlinear responses, having a broad span of potential applications by themselves[11], still pose challenges to attempts of employing them as building blocks for the emulation of complex nonlinear dynamics. On the other hand, the RF technology offers mature fabrication and characterization techniques, suggesting a more amenable scenario for investigating nonlinear elements in this spectral range. In particular, higher-harmonic generation in various designs, such as varactor diodes embedded in RF circuits, has been demonstrated[12]. Further control of nonlinear responses in coupled- resonator geometries was achieved via tuning their relative positions[13].

In this Letter, the doubly-resonant coupled split-rings resonator's (SRR) geometry is employed for achieving tuning and enhancement of nonlinear responses (Fig. 1). It is shown that the structure may serve as a building block for creating nonlinear meta-atoms with nonlinearity on demand. The Letter is organized as follows: first, the coupled-rings design is elaborated. Then two types of nonlinear experiments are performed – the first one is near-to-near field generation, where both the pump and nonlinear harmonics are excited/collected by means of broadband loop probes. The second experiment is aimed at the near-to-far field scenario, where nonlinear-wave properties are scanned at a distance. The consideration is focused on the second- and third harmonic generation (SHG and THG), and approaches for controlling relations between them. The developed method can be straightforwardly applied to control of higher-order harmonics as well.

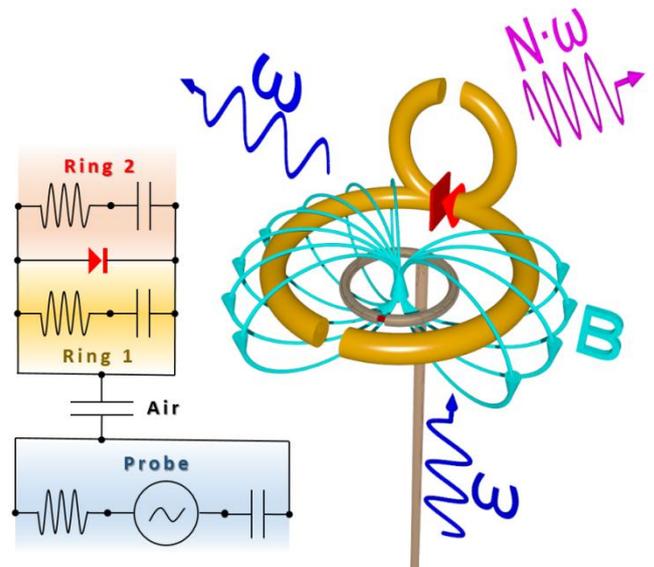

*Fig 1. (Color Online) Schematics of the doubly-resonant nonlinear meta-atom, built as a pair of split-ring resonators sharing a varactor diode at a joint gap. The grey loop underneath the lower ring is an excitation probe. The left inset displays a lumped circuit scheme of the system.*

**Results**
*The general concept of the double-ring* structure employed here is presented in Fig. 1. It contains a pair of coupled SRRs[14], with the first (larger) ring acting as a receiver with the resonance tuned to the incident-wave carrier frequency (ω). The second ring is a transceiver, and its resonance is adjusted to fit a higher harmonic, i.e., 2ω or 3ω in the present setting. The source of the nonlinearity is a varactor diode, which is placed at a joint gap, shared by both SRRs. A voltage drop on the diode produces a current which contains multiple harmonics, due to its nearly exponential I-V response. To enable radiating those harmonics to the far field, a matching circuit should be introduced, which is the primary function of the second ring. This element also enables filtering nonlinear harmonics, by enhancing a single one and inhibiting all the others (e.g., the suppressing the third harmonic is favor of the second, or vice versa, depending on to the purpose).

The lumped circuit scheme, representing the nonlinear interaction, is rendered in the left inset to Fig. 1. The equivalent circuit of each individual ring contains capacitive, inductive and resistive elements (the latter is not shown in the scheme), whose strengths can be directly controlled by adjusting the geometry, viz., the size of the gaps and radii of the rings. The spatial orientation of the rings is chosen to be mutually perpendicular, to reduce the linear coupling between them. In this case, the pump wave excites the receiver and the nonlinear current, induced in the varactor, radiates the far field via the transmitter. Note that the single nonlinear element in the meta-atom is an optical counterpart of an anharmonic intrinsic potential in natural atoms or molecules. Furthermore, the relative spatial orientation of the rings enables controlling components of the nonlinear susceptibility tensor, as a linearly exited dipole is not necessarily aligned with the resulting effective nonlinear source. In principle, varying relative spatial orientations of the rings (and probably adding others) will enable achieving a nonlinear tensor with pre-defined matrix elements.

*Experimental samples of the double rings* are based on 0.1 mm thick copper strips, chemically etched on an FR4 substrate. While the mutual linear coupling between the rings was minimized by setting them perpendicular to each other, it still may affect original resonance characteristics obtained for standalone geometries. Numerical optimization, performed by dint of the finite-elements method (CST Microwave Studio), enables designing the structure with the parasitic capacitance between the rings and the measurement apparatus (loop probes) taken into account. Substrate thicknesses and their tolerances, which have a significant impact on the design, were also taken into account. Initial parameters for the optimization were chosen as those of isolated rings, resonating at pump and nonlinear frequencies, accordingly. The double-rings structures were optimized to feature the doubly-resonant properties. Numerical values of reflection amplitudes (absolute values of $S_{11}$ parameter) of the designed structures appear in Fig. 2, where the double-peak behavior is clearly observed. Furthermore, the structure, resonating at $2\omega$, filters out the signal at $3\omega$ (the blue line and inset (a)), while the resonance tuned to the third harmonic inhibits the second one (the red line and inset (b)). On the other hand, adjusting parameters of the setup, it is also possible to realize a situation with simultaneous mutually coherent emission of the second and third harmonics, which is the case of considerable interest too[15].

Table 1 summarizes geometric properties of the SRRs that appear in insets to Fig. 2. A SMV1232 varactor diode[16], with DC-voltage controllable capacitance in the range of $1-4\ pF$ was used, and its properties were included in the simulations by incorporating the corresponding lumped circuit element.

| Ring | Radius[$mm$] | Gap width [$mm$] |
|---|---|---|
| $\omega$ | 9.68 | 1 |
| $2\omega$ | 4.98 | 1 |
| $3\omega$ | 3.5 | 1 |

*Table 1. Parameters on the fabricated SRRs.*

*Near-to-near field generation.* To verify the performance of the samples, both excitation and collection efficiencies were maximized, making use of coupling to near-field broadband loop probes. They were connected to coaxial ports of a vector network analyzer (Agilent E8362B), which is able to extract both the amplitude and phase of the received signals. The SHG efficiency in the double-rings geometry was compared to the performance of the single ring. The loop probe beneath the larger ring (Fig. 1) excited the sample at $1.737\ GHz$, while the second probe collected the transmitted spectrum. Figure 3 compares the SHG efficiency of the single and double-ring samples. While in both cases the power of the second harmonic depends linearly on the pump power (as expected from the second-order nonlinear processes), slopes ($\eta$) indicate different efficiencies of the samples. Experimental results show $\eta_{single} = 0.23$, $\eta_{double} = 0.43\ \pm 0.1\%$, i.e., the doubly-resonant structure outperforms the single ring by a factor of ~2. The nonlinear signal is measured as a voltage drop (in mV) acquired by the VNA, the units of the horizontal axis being mW. It should be noted, however, that absolute values of the nonlinear efficiency cannot be easily mapped into intrinsic efficiency of the device, as the probe-to-device coupling strongly depends on the relative position between them. Imperfections in matching circuits also affect the overall efficiency produced by the experiment.

To verify the resonant dependence of the SHG on the pump frequency, the efficiency of the process in the double-ring geometry was measured as a function of the pump frequency (the slope in Fig. 3 contributes one point to this dependence). The inset to Fig. 3 summarizes those results, underlining a clearly resonant behavior, peaked around the resonance of the first ring ($1.737\ GHz$). The experimental data was fit to a Lorentzian profile, with a full width at half maximum of 7.4 KHz.

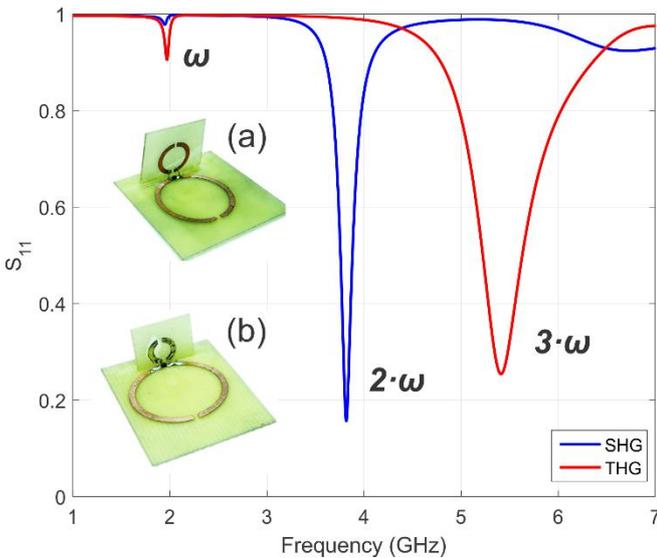

*Fig 2. (Color Online). Reflection amplitudes (the absolute values of the $S_{11}$ parameters) of the double-rings setting (numerical data), tuned for second-harmonic or third-harmonic generation (blue and red lines, respectively). Insets are photographs of the fabricated samples: (a) for SHG, (b) for THG.*

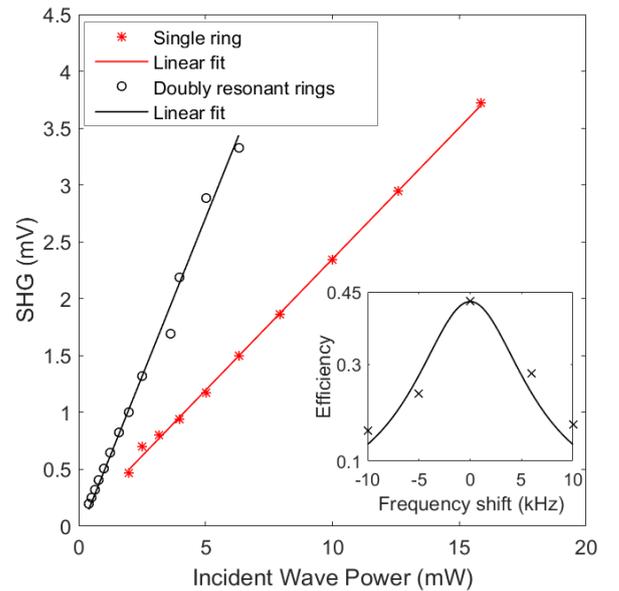

*Fig 3. (Color Online). The second-harmonic generation efficiency in the near-to-near field configuration, as a function of the incident power, at $f = 1.737\ GHz$. Dots – experimental data, lines – a linear fit to them. The insert shows the SHG efficiency as a function of the pump-frequency detuning (in KHz) from the central resonant frequency.*

*Controlling harmonics by means of auxiliary structures in the framework of the near-to-near field generation.* Having established the resonant nature of the nonlinear enhancement effect, it can be controlled with the help of the auxiliary geometries. The varactor diode, having a highly nonlinear I-V response, generates the entire comb of nonlinear harmonics[12]. While the intrinsic generation efficiency of those harmonics depends on the internal structure of the varactor, auxiliary material structures can either enhance or suppress both near- and far-field signatures of the nonlinear microscopic currents by performing linear filtering. The filtering properties can be understood by recalling the transmission data, which is displayed in Fig. 2.

To demonstrate the on-demand control of nonlinear properties by means of auxiliary structures, two samples were compared, with the smaller ring optimized either to the SHG or THG generation (insets to Fig. 2). It is worth noting that the intrinsic high-order nonlinearities of the varactor are comparable[12], in contrast to the usual behavior of optical materials. Figure 4 summarizes the results for the near-to-near field transmission spectrum. It is seen that the auxiliary second ring completely suppresses undesired harmonics. Figure 4(a) shows efficient generation of the second harmonic and complete suppression of the third. The picture is completely changed by constructing the second ring to resonate at the third harmonic: in this case, THG is highly efficient, while SHG is suppressed (Fig. 4(b)). It should be stressed again that the auxiliary ring acts as a linear filter and does not operate at the microscopic (PN junction of the varactor) level, effectively changing the intrinsic nonlinearity of the varactor.

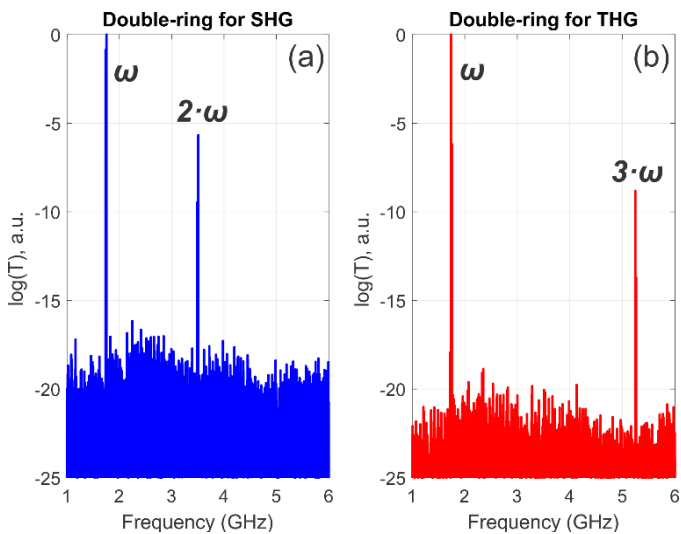

*Fig 4. (Color Online). The near-to-near field transmission (experimental data on the logarithmic scale) for the double ring optimized for SHG (a), and THG (b). In both cases, the transmission at the pump frequency is normalized to unity.*

*Controlling harmonics by auxiliary structures in the framework of near-to-far field generation.* To verify the far-field characteristics of the emission from the nonlinear meta-atoms, spectra were recorded, using an anechoic chamber. While the pump field was launched via the same near-field probe, the far field was collected by a horn antenna, positioned on the same axis, perpendicular to the direction of the nonlinear dipole (the direction of the maximum radiation power). The distance between the nonlinear atom and horn was 1.5m, ensuring the far-field operation. Figure 5 summarizes the results. The enhancement of the selected harmonic and suppression of the other is observed, similarly to the near-to-near filed case, considered above. Relative ratios between the SHG and THG intensities in the near- and far-field configurations are slightly different, due to dissimilar coupling efficiencies of the collection probes (the loop and horn in the near-field far-field settings, respectively).

The overall nonlinear efficiency of the devices can be investigated numerically by means of so called two-step simulation[17]. Namely, the pump-field distribution is calculated at the first stage, serving to estimate the corresponding nonlinear sources in terms of nonlinear susceptibilities. At the second step, the linear propagation from the new sources, oscillating at a higher frequency, is simulated. This approach assumes the undepleted pump, and makes it possible to perform two linear simulations, instead of solving the highly nonlinear sophisticated problem[9]. To investigate the field propagator at the second stage, the far-field characteristics of the emitting nonlinear meta-atoms were computed numerically, see the results appear in insets to Fig. 5. In both SHG and THG cases, a point dipole, emulating the nonlinear current in the diode, was used as the source. Although the radiation patterns were affected by the double-rings structures, in both cases the patterns are similar to those produced by isolated dipoles. Radiation efficiencies of the dipoles are enhanced by the auxiliary rings, selectively matching impedances to the far-field.

Knowing the dipolar patterns of nonlinear fields generated by the meta-atoms, one may proceed direct emulation of nonlinear materials by assembling the atoms in ordered one-, two-, and three-dimensional arrays. Moreover, by imprinting phase-shift patterns onto the arrays, it is possible to produce topological radiation modes, such as vortices, or even skyrmions and hopfions.

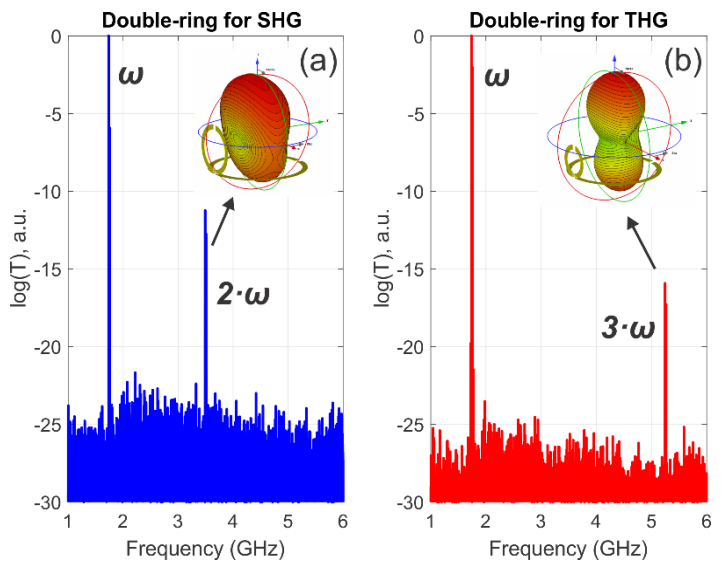

*Fig 5. (Color Online). The near-to-far field transmission (experimental data on the logarithmic scale) for the double-ring optimized for SHG (a), and THG (b). In both cases, the transmission at the pump frequency is normalized to unity. Insets display far-field radiation patterns.*

**Conclusion** The concept of doubly-resonant meta-atoms with nonlinear responses, tailored on demand, is proposed and demonstrated in the GHz spectral range. In particular, it is shown that power shares carried by the second and third harmonics can be tailored, beyond intrinsic properties of the nonlinear source, by means of material structures built around it. The concept can be extended for tailoring other nonlinear properties, such as combined generation of higher harmonics, by means of the accurate design of the control structures. While the similarities in the nonlinear interactions of electromagnetic waves with natural and artificial materials are valid to a certain extent, the latter may serve as an element of a large-scale toolbox for the emulation of complex nonlinear phenomena.

**Acknowledgments** This work was supported, in part, by grants from the Fund of the Tel Aviv University Rector, German-Israel Foundation (grant number 2399), the Government of the Russian Federation (grant No. 074-U01), and Russian Fund for Basic Research (project No. 16-52-00112).

**References:**
[1] Y.S. Kivshar and G. Agrawal, *Optical Solitons: From Fibers to Photonic Crystals* (Academic Press; 1 edition, 2003).
[2] T.G. Philbin, C. Kuklewicz, S. Robertson, S. Hill, F. König, and U.

Leonhardt, Science **319**, 1367 (2008).
[3] N. Engheta and R.W. Ziolkowski, *Electromagnetic Metamaterials: Physics and Engineering Explorations* (Wiley-IEEE Press; 1 edition, 2006).
[4] M. Lapine, I. V. Shadrivov, and Y.S. Kivshar, Rev. Mod. Phys. **86**, 1093 (2014).
[5] M. Kauranen and A. V. Zayats, Nat. Photonics **6**, 737 (2012).
[6] A.P. Slobozhanyuk, P. V. Kapitanova, D.S. Filonov, D.A. Powell, I. V. Shadrivov, M. Lapine, P.A. Belov, R.C. McPhedran, and Y.S. Kivshar, Appl. Phys. Lett. **104**, 1 (2014).
[7] P. Ginzburg, A. Krasavin, Y. Sonnefraud, A. Murphy, R.J. Pollard, S.A. Maier, and A. V. Zayats, Phys. Rev. B **86**, 085422 (2012).
[8] P. Ginzburg, A. V Krasavin, G.A. Wurtz, and A. V Zayats, ACS Photonics 141212092238000 (2014).
[9] A. V Krasavin, P. Ginzburg, G.A. Wurtz, and A. V Zayats, Nat. Commun. **7**, 11497 (2016).
[10] Z.Y. Ou and Y.J. Lu, Phys. Rev. Lett. **83**, 2556 (1999).
[11] N. Segal, S. Keren-Zur, N. Hendler, and T. Ellenbogen, Nat. Photonics **9**, 180 (2015).
[12] I. V. Shadrivov, A.B. Kozyrev, D.W. van der Weide, and Y.S. Kivshar, Appl. Phys. Lett. **93**, 161903 (2008).
[13] K.E. Hannam, D.A. Powell, I. V. Shadrivov, and Y.S. Kivshar, Appl. Phys. Lett. **100**, 081111 (2012).
[14] A. Velez, J. Bonache, and F. Martin, IEEE Microw. Wirel. Components Lett. **18**, 28 (2008).
[15] O. Bang, J. Opt. Soc. Am. B **14**, 51 (1997).
[16] Skyworks, (n.d.).
[17] P. Segovia, G. Marino, A. V. Krasavin, N. Olivier, G.A. Wurtz, P.A. Belov, P. Ginzburg, and A. V. Zayats, Opt. Express **23**, 30730 (2015).